\documentclass[twocolumn,prl,aps,nobalancelastpage]{revtex4}

\usepackage{bm}

\usepackage{graphicx}

\begin{document}

\title{Magnetic penetration depth of single crystal SmFeAsO$_{1-x}$F$_y$: a fully gapped superconducting state}

\author{L. Malone, J.D. Fletcher, A. Serafin, and A. Carrington}

\affiliation{H.\,H. Wills Physics Laboratory, University of Bristol, Tyndall Avenue, BS8 1TL, United Kingdom.}

 \author{N.D. Zhigadlo, Z. Bukowski,  S. Katrych, and J. Karpinski}

\affiliation{Laboratory for Solid State Physics, ETH, 8093 Z\"urich, Switzerland.}

\date{\today}
\begin{abstract}
We report measurements of the in-plane magnetic penetration depth $\lambda$ in single crystals of SmFeAsO$_{1-x}$F$_y$
($x\simeq y \simeq 0.2$) with $T_c \simeq 45$\,K.  We find that at low temperature $\lambda$ has an exponential
temperature dependence which suggests that the Fermi surface is fully gapped.  The magnitude of the minimum energy gap,
$\Delta_1=1.1\pm 0.1 k_BT_c$ at $T=0$\,K, is significantly smaller than the BCS weak-coupling value suggesting that the
gap is either strongly anisotropic or varies significantly between the different Fermi surface sheets. Our data fits
well a two gap model with an additional larger gap of magnitude $\Delta_2 = 1.8\pm0.2 k_BT_c$ which is associated with
$\sim 80$\% of the total superfluid density.
\end{abstract}

\maketitle

Superconductivity in the Fe oxypnictide compounds, LnFeAsO$_{1-x}$F$_y$ (where Ln = La, Sm, Ce, Nd, or Pr) has
generated an enormous amount of interest.  The maximum value of $T_c\simeq 55$\,K \cite{Ren2008} found so far in this
series, is the highest for any non-cuprate superconductor.  It is significantly higher than that found for the previous
non-cuprate record holder MgB$_2$ ($T_c\simeq$ 40\,K). The electronic structure of these materials has many
similarities with the cuprates. Calculations have shown that the Fermi surface is expected to be quasi-two-dimensional
and strong ferromagnetic and antiferromagnetic spin-fluctuations are predicted \cite{SinghPRL08,IIMazin08032740}.
Importantly, the calculations suggest that the electron-phonon interactions are much too weak to produce a $T_c$ of
$\sim$ 55\,K \cite{IIMazin08032740}.  Hence, in many ways it might be expected that the superconductivity has more in
common with the cuprates than the phonon mediated superconductor MgB$_2$.

The determination of the symmetry of the superconducting order parameter is an important first step toward uncovering
the mechanism of superconductivity in any material.  In this regard, measurements of the magnetic penetration depth
$\lambda$ have played an important role.  Although not a true bulk probe, like specific heat, penetration depth
measurements in the Meissner state probe a few thousand Angstroms below the crystal surface and so should be reasonably
representative of the bulk.  However, experience with cuprates and MgB$_2$ has shown that the data are most reliable if
measurements are performed on high quality single crystal samples --- particularly for strongly anisotropic materials
such as the pnictides \cite{SinghPRL08,IIMazin08032740,SWeyeneth08061024}.

There have been several reports of experiments to determine the superconducting gap structure in these pnictide
materials. In particular, several groups have reported the results of point contact Andreev spectroscopy measurements
(PCAS). Chien \emph{et al.} \cite{TYChenNature08} report data for polycrystalline SmFeAsO$_{0.85}$F$_{0.15}$
($T_c$=40\,K) which can be fitted by a single isotropic ($s$-wave) gap $\Delta_0/k_BT_c = 1.9$ which follows well the
BCS weak-coupling temperature dependence. However, Wang \emph{et al.} \cite{YongleiWang08061986} report PCAS data also
on
 SmFeAsO$_{1-x}$F$_{x}$ with $T_c=51.5$\,K which is best described by a $d$-wave gap symmetry with two distinct gap
magnitudes ($\Delta_0/k_BT_c$= 1.1 and 2.9). The data show a pronounced zero bias anomaly which is interpreted as
further evidence for nodes in the gap function. Zero bias anomalies have also been reported in PCAS data in the Nd
compound \cite{PSamuely08061672}. It is likely these differences are reflective of the state of the surfaces of the
samples, and further measurements are needed on high quality single crystals before the issue can be settled.

In this paper, we report measurement of the in-plane magnetic penetration depth of single crystal SmFeAsO$_{1-x}$F$_y$
($x\simeq y \simeq 0.2$) with $T_c \simeq 45$\,K.  The data below $T\sim 15$\,K show that $\lambda$ follows an
exponential temperature dependence, characteristic of an $s$ wave, fully gapped superconductor.  The value of the
minimum energy gap is found to be significantly lower than the BCS weak-coupling value.

Single crystals of nominal composition SmFeAsO$_{0.8}$F$_{0.2}$ were grown in Z\"urich using a high-pressure cubic
anvil technique and a NaCl/KCl flux \cite{NDZhigadlo08060337}.  The small plate-like single crystals have typical
dimensions (80$\times$80$\times$20$)\mu$m$^3$, with the smallest dimension being along the $c$-axis.  The penetration
depth was measured using a radio frequency ($F\simeq$ 12 MHz) tunnel diode oscillator technique \cite{CarringtonGKG99}.
The sample was attached with vacuum grease to the end of a high purity sapphire rod and placed in a solenoid coil,
which forms part of the inductor of the resonant circuit. The magnetic field was directed along the samples $c$
direction and hence all the screening currents flow in the $ab$ plane. We estimate that the RF field was $\sim 10^{-7}$
T and the Earth's field was screened with a mu-metal can, hence, we do not expect any contributions from mobile
vortices.  The change in the resonant frequency of the circuit as the temperature is varied is directly proportional to
the change in $\lambda_{ab}$. The constant of proportionality was determined from the measured ab-plane dimensions of
the samples and the total frequency shift obtained when the sample was extracted from the coil at base temperature
\cite{ProzorovGCA00}. We estimate that this is accurate to $\sim$ 20\%.

\begin{figure}
\includegraphics[width=8.0cm,keepaspectratio=true]{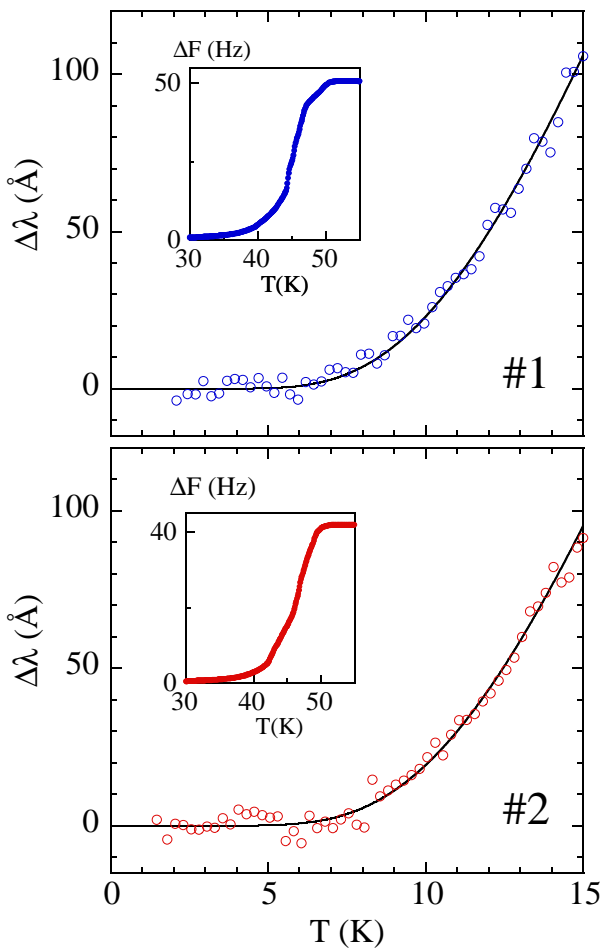}
 \caption{(color online). Temperature dependence of the in-plane penetration depth of single
 crystals of SmFeAsO$_{0.8}$F$_{0.2}$.  The two panels show data for two different crystals. The
 solid lines are fits to Eq.\ (\ref{eqbcs}).  The insets show the raw data for the frequency shifts close to the
 superconducting transition.}
 \label{figlowT}
\end{figure}

Results for the low temperature behavior of two different samples are shown in Fig.\ \ref{figlowT}.  Here we denote the
change in $\lambda_{ab}$ from the values at our lowest measurement temperature as $\Delta \lambda(T)$.  The results for
both samples are very similar, showing that below $T\simeq 6$\,K the penetration depth becomes independent of
temperature within the noise of our measurements.  The absolute change of $\lambda$ between 2\,K and 15\,K was also
quite consistent, being 105 \AA~ for sample \#1 and 95~\AA~for sample \#2. This small difference is well within our
expected error.

In the BCS theory for a fully gapped $s$-wave superconductor, the behavior of $\lambda(T)$ asymptotically approaches
\begin{equation}
\Delta \lambda (T) = \lambda(0) \sqrt{\frac{\pi\Delta_0}{2k_BT}}\exp\left(-\frac{\Delta_0}{k_BT}\right) \label{eqbcs}
\end{equation}
at low temperature.  Here $\lambda(0)$ and $\Delta_0$ are the values of $\lambda$ and the superconducting energy gap
$\Delta$ at $T=0$.  In practice this provides a good approximation to the full theory for $T \lesssim T_c/3$. If the
energy gap is (weakly) anisotropic or there are distinct gaps on different Fermi surface sheets then the same behavior
will still be found but with $\Delta_0$ now being approximately equal to the minimum energy gap in the system, and
$\lambda(0)$ is replaced by an effective value which depends on the details of the gap anisotropy.

The solid lines in Fig.\ \ref{figlowT} show fits of the data for samples \#1 and \#2 to Eq.\ (\ref{eqbcs}) giving
values of $\Delta_0/k_B=49\pm 2$\,K and $55\pm5$\,K respectively (4.2meV and 4.7meV).  The fitted values of $\Delta_0$
depend slightly on the upper temperature limit of the data included in the fit, and the quoted error bars encompassed
the spread in values obtained for an upper temperature limit less than 15\,K ($\simeq T_c/3$).  The behavior of the
susceptibility close to the superconducting transition for these two samples is shown in the inset to Fig.\
\ref{figlowT}. Although there are differences between the two samples in the structure near $T_c$, the similarity of
the low temperature results suggests that they are not affected by the slight inhomogeneity present. The onset of the
transition is at around 50\,K with the midpoint at around 45\,K. From this we estimate that $\Delta_0/k_BT_c =
1.1\pm0.1$, which is significantly lower than the weak-coupling $s$-wave BCS value of 1.76, and suggests the
possibility of significant gap anisotropy or multiple gaps such as is found in, for example, NbSe$_2$
\cite{fletcher2007} and MgB$_2$ \cite{ManzanoCHLYT02}.  The value of $\Delta_0/k_BT_c$ found here is significantly
higher than that found for the small gap in MgB$_2$ ($\Delta_0/k_BT_c=0.76$), and is closer to the value found for the
minimum gap in NbSe$_2$ \cite{fletcher2007}. Evidence for multi-gap behavior in SmFeAsO$_{1-x}$F$_y$ has also been
inferred from strong observed temperature dependence of the anisotropy parameter $\gamma$ \cite{SWeyeneth08061024}.

Given the many similarities with the cuprates, there has naturally been much speculation regarding the possibility that
the superconductivity in these pnictide materials is unconventional. For a simple $d$-wave superconductor at low
temperature, $\Delta \lambda \simeq \ln 2 \lambda(0) k_B T/\Delta_0$. Assuming values appropriate for a $d$-wave state
in these materials ($\lambda(0)=2000$\AA~and $\Delta_0=2.14~ k_BT_c$) we estimate that $d\lambda/dT \simeq 14$ \AA/K
which is at least two orders of magnitude larger than any linear term present in our data below $\sim 6$\,K. In a
gapless superconductor, impurities produce a finite zero energy density of states and the temperature dependence of
$\lambda$ changes smoothly from $T$ to $T^2$, below an temperature scale determined by the impurity concentration
\cite{HirschfeldG93}.  This is also incompatible with our data, however, Cooper \cite{Cooper96} has shown that if the
normal state of a superconductor is strongly paramagnetic then there is an additional contribution to the measured
$\lambda$ which is approximately equal to $\lambda(0)\chi_N(T)/2$, where $\chi_N$ is the normal state susceptibility.
This additional term can produce a minimum in the measured $\lambda(T)$ which over a short range of temperature which
can resemble an exponential BCS-like $T$ dependence (see for example the measurements on the cuprate superconductor
Nd$_{1.85}$Ce$_{0.15}$CuO$_{4-y}$ which are discussed and reanalysed in Ref.\ \cite{Cooper96}). We have considered this
possibility, assuming that $\chi_N = C/(T+T_M)$, but were unable to obtain a reasonable fit to our data, even when we
allowed all the parameters to vary freely. We conclude that a gapless superconducting state appears to be ruled out by
our data.

Electronic structure calculations for the sister material LaFeAsO$_{1-x}$F$_x$ show that the bands crossing the Fermi
level originate mostly from the two dimensional Fe layer and give rise to quasi-two-dimensional cylindrical sheets
running along the c-axis and centered on the $\Gamma$ point (hole) and the $M$ point (electron). The general topology
of this Fermi surface is consistent with recent angle resolved photoemission measurements \cite{CLiu08062147}.  As
discussed by Mazin \emph{et al.}\cite{IIMazin08032740} this Fermi surface topology places constraints on the type of
pairing symmetries which are likely.  In particular, these authors argue that the structure of the magnetic
fluctuations favors an unusual type of multiple gap $s$-wave superconductivity where the order parameter on the
electron and hole Fermi surface sheets have opposite phase.  Our experiment does not tell us anything about the phase
structure but our results are compatible with a fully gapped $s$-wave state with moderate anisotropy. Also, the
presence of two different types of Fermi surface sheets suggests the possibility of two-gap superconductivity although
there is expected to be significant intersheet scattering which would make the magnitudes of the two gaps similar
\cite{IIMazin08032740}.

\begin{figure}
\centerline{\includegraphics[width=8.0cm,keepaspectratio=true]{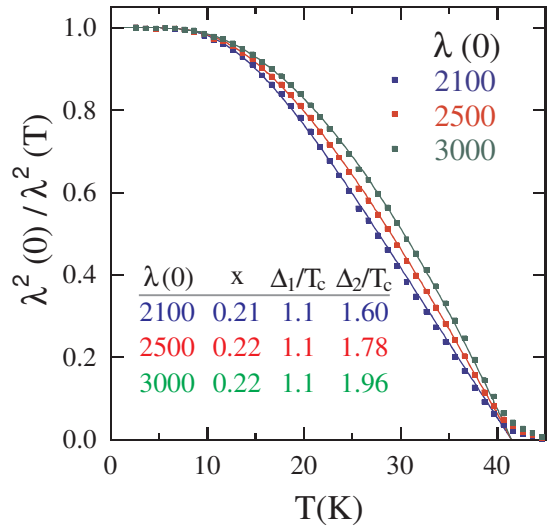}}
 \caption{(color online).  Calculated normalized superfluid density [$\lambda^2(0)/\lambda^2(T)$] versus temperature for sample \#1, using different assumed
 values for the zero temperature penetration depth $\lambda(0)$. The solid lines are fits to the two gap model described in the text, and the inset table
 shows the fit parameters.}
 \label{figrho}
\end{figure}

As remarked above, fitting the low temperature $\lambda(T)$ data to Eq.\ (\ref{eqbcs}) will deduce the size of the
minimum superconducting gap.  To deduce the presence of any other gaps or any other form of gap anisotropy it is
necessary to analyze the full temperature dependence of the normalized superfluid density
$\rho=\lambda^2(0)/\lambda^2(T)$ up to $T_c$. In our experiment we are not able to measure the absolute value of
$\lambda$. So in order to calculate the superfluid density it is necessary to use an estimate of $\lambda(0)$ from
other experiments. Weyeneth \emph{et al.} \cite{SWeyeneth08061024} report measurement of $\lambda$ deduced from the
reversible magnetic torque measured on samples produced using exactly the same method as those in the present work with
a similar $T_c$. From their data they estimate $\lambda_{ab}(0)$=2100 $\pm$ 300\,\AA.  This is in reasonable agreement
with the value of $\lambda\simeq $ 1900\,\AA~measured by $\mu$SR for SmFeAsO$_{0.85}$ ($T_c=52$\,K)
\cite{Khasanov08051923}.

In Fig.\ \ref{figrho} we plot the superfluid density calculated from our data using a range of values of $\lambda(0)$
that approximately encompass \cite{lam0note} both the expected uncertainty in both $\lambda(0)$ and our calibration
factor relating the measured frequency shifts to $\Delta \lambda (T)$. We fit this superfluid density with a two-gap
BCS model as used for MgB$_2$ \cite{ManzanoCHLYT02,Fletcher2005}. Here it is assumed that the gaps ($\Delta_1$ and
$\Delta_2$) on each Fermi surface sheet (or group of sheets) follow the weak-coupling BCS temperature dependence but
have a variable low temperature absolute value. There is an additional parameter $x$ which is the fraction of the
superfluid density on the sheet with the smaller gap. In principle, there are four free parameters,($x$, $T_c$,
$\Delta_1$ and $\Delta_2$) but we fix $\Delta_1=1.1\,T_c$ as suggested by the low temperature exponential fits, and
$T_c=41.5$\,K as suggested by a linear extrapolation of the superfluid density close to $T_c$. The fits suggests that
in addition to the small gap found from the analysis of the low temperature data there is a larger gap of value
$\Delta_2=1.8\pm 0.2$ ($6.4\pm0.7$ meV) which accounts for $80\pm5$\% of the superfluid density. This value of
$\Delta_2$ is close to the gap value measured by Chien \emph{et al.} \cite{TYChenNature08} using PCAS on
SmFeAsO$_{0.85}$F$_{0.15}$ ($T_c$=40\,K) mentioned above. The small amount of superfluid density associated with the
smaller gap may explain why this was not observed in these PCAS measurements. We note that we can also fit our data to
a model where there is a smooth variation of the gap with in-plane angle [i.e., $\Delta (\phi)=\Delta (1+ \varepsilon
\cos 4 \phi)$], so our data does not give definitive evidence for two gaps but rather just for a moderate variation of
$\Delta(\bm{k})$.

In summary, our data for the in-plane penetration depth of the Sm Fe oxypnictide superconductor
SmFeAsO$_{0.8}$F$_{0.2}$ show an exponential temperature dependence at low temperature indicating a fully gapped
pairing state.  Our results also show evidence for a moderate variation of the gap on the different Fermi surface
sheets.  Fitting our data with a two-gap model suggest that the second gap is $\sim 60$\% larger than the first.

We thank J.R.\ Cooper and I.I.\ Mazin for useful discussions. This work was supported by the UK EPSRC, and the Swiss
National Science Foundation through the NCCR pool MaNEP.

%\bibliography{SmFeLambdaT}

\begin{thebibliography}{17}
\expandafter\ifx\csname natexlab\endcsname\relax\def\natexlab#1{#1}\fi \expandafter\ifx\csname
bibnamefont\endcsname\relax
  \def\bibnamefont#1{#1}\fi
\expandafter\ifx\csname bibfnamefont\endcsname\relax
  \def\bibfnamefont#1{#1}\fi
\expandafter\ifx\csname citenamefont\endcsname\relax
  \def\citenamefont#1{#1}\fi
\expandafter\ifx\csname url\endcsname\relax
  \def\url#1{\texttt{#1}}\fi
\expandafter\ifx\csname urlprefix\endcsname\relax\def\urlprefix{URL }\fi \providecommand{\bibinfo}[2]{#2}
\providecommand{\eprint}[2][]{\url{#2}}

\bibitem[{\citenamefont{Ren et~al.}(2008)\citenamefont{Ren, Lu, Yang, Yi, Shen,
  Li, Che, Dong, Sun, Zhou, and Zhao}}]{Ren2008}
\bibinfo{author}{\bibfnamefont{Z.}~\bibnamefont{Ren}},
  \bibinfo{author}{\bibfnamefont{W.}~\bibnamefont{Lu}},
  \bibinfo{author}{\bibfnamefont{J.}~\bibnamefont{Yang}},
  \bibinfo{author}{\bibfnamefont{W.}~\bibnamefont{Yi}},
  \bibinfo{author}{\bibfnamefont{X.-L.} \bibnamefont{Shen}},
  \bibinfo{author}{\bibfnamefont{Z.-C.} \bibnamefont{Li}},
  \bibinfo{author}{\bibfnamefont{G.-C.} \bibnamefont{Che}},
  \bibinfo{author}{\bibfnamefont{X.-L.} \bibnamefont{Dong}},
  \bibinfo{author}{\bibfnamefont{L.-L.} \bibnamefont{Sun}},
  \bibinfo{author}{\bibfnamefont{F.}~\bibnamefont{Zhou}}, \bibnamefont{and}
  \bibinfo{author}{\bibfnamefont{Z.-X.} \bibnamefont{Zhao}},
  \bibinfo{journal}{Chin. Phys. Lett.} \textbf{\bibinfo{volume}{25}}
  (\bibinfo{year}{2008}).

\bibitem[{\citenamefont{Singh and Du}(2008)}]{SinghPRL08}
\bibinfo{author}{\bibfnamefont{D.~J.} \bibnamefont{Singh}} \bibnamefont{and}
  \bibinfo{author}{\bibfnamefont{M.-H.} \bibnamefont{Du}},
  \bibinfo{journal}{Phys. Rev. Lett.} \textbf{\bibinfo{volume}{100}},
  \bibinfo{pages}{237003} (\bibinfo{year}{2008}).

\bibitem[{\citenamefont{Mazin et~al.}(2008)\citenamefont{Mazin, Singh,
  Johannes, and Du}}]{IIMazin08032740}
\bibinfo{author}{\bibfnamefont{I.}~\bibnamefont{Mazin}},
  \bibinfo{author}{\bibfnamefont{D.}~\bibnamefont{Singh}},
  \bibinfo{author}{\bibfnamefont{M.}~\bibnamefont{Johannes}}, \bibnamefont{and}
  \bibinfo{author}{\bibfnamefont{M.}~\bibnamefont{Du}},
  \bibinfo{journal}{arXiv:0803.2740 (unpublished)}  (\bibinfo{year}{2008}).

\bibitem[{\citenamefont{Weyeneth et~al.}()\citenamefont{Weyeneth, Mosele,
  Zhigadlo, Katrych, Bukowski, Karpinski, Kohout, Roos, and
  Keller}}]{SWeyeneth08061024}
\bibinfo{author}{\bibfnamefont{S.}~\bibnamefont{Weyeneth}},
  \bibinfo{author}{\bibfnamefont{U.}~\bibnamefont{Mosele}},
  \bibinfo{author}{\bibfnamefont{N.}~\bibnamefont{Zhigadlo}},
  \bibinfo{author}{\bibfnamefont{S.}~\bibnamefont{Katrych}},
  \bibinfo{author}{\bibfnamefont{Z.}~\bibnamefont{Bukowski}},
  \bibinfo{author}{\bibfnamefont{J.}~\bibnamefont{Karpinski}},
  \bibinfo{author}{\bibfnamefont{S.}~\bibnamefont{Kohout}},
  \bibinfo{author}{\bibfnamefont{J.}~\bibnamefont{Roos}}, \bibnamefont{and}
  \bibinfo{author}{\bibfnamefont{H.}~\bibnamefont{Keller}},
  \bibinfo{journal}{arXiv:0806.1024 (unpublished)}.

\bibitem[{\citenamefont{Chen et~al.}(2008)\citenamefont{Chen, Tesanovic, Liu,
  Chen, and Chien}}]{TYChenNature08}
\bibinfo{author}{\bibfnamefont{T.~Y.} \bibnamefont{Chen}},
  \bibinfo{author}{\bibfnamefont{Z.}~\bibnamefont{Tesanovic}},
  \bibinfo{author}{\bibfnamefont{R.~H.} \bibnamefont{Liu}},
  \bibinfo{author}{\bibfnamefont{X.~H.} \bibnamefont{Chen}}, \bibnamefont{and}
  \bibinfo{author}{\bibfnamefont{C.~L.} \bibnamefont{Chien}},
  \bibinfo{journal}{To appear in Nature}.

\bibitem[{\citenamefont{Wang et~al.}()\citenamefont{Wang, Shan, Fang, Cheng,
  Ren, and Wen}}]{YongleiWang08061986}
\bibinfo{author}{\bibfnamefont{Y.}~\bibnamefont{Wang}},
  \bibinfo{author}{\bibfnamefont{L.}~\bibnamefont{Shan}},
  \bibinfo{author}{\bibfnamefont{L.}~\bibnamefont{Fang}},
  \bibinfo{author}{\bibfnamefont{P.}~\bibnamefont{Cheng}},
  \bibinfo{author}{\bibfnamefont{C.}~\bibnamefont{Ren}}, \bibnamefont{and}
  \bibinfo{author}{\bibfnamefont{H.-H.} \bibnamefont{Wen}},
  \bibinfo{journal}{arXiv:0806.1986 (unpublished)}.

\bibitem[{\citenamefont{Samuely et~al.}()\citenamefont{Samuely, Szabo,
  Pribulova, Tillman, Bud'ko, and Canfield}}]{PSamuely08061672}
\bibinfo{author}{\bibfnamefont{P.}~\bibnamefont{Samuely}},
  \bibinfo{author}{\bibfnamefont{P.}~\bibnamefont{Szabo}},
  \bibinfo{author}{\bibfnamefont{Z.}~\bibnamefont{Pribulova}},
  \bibinfo{author}{\bibfnamefont{M.~E.} \bibnamefont{Tillman}},
  \bibinfo{author}{\bibfnamefont{S.}~\bibnamefont{Bud'ko}}, \bibnamefont{and}
  \bibinfo{author}{\bibfnamefont{P.~C.} \bibnamefont{Canfield}},
  \bibinfo{journal}{arXiv:0806.1672 (unpublished)}.

\bibitem[{\citenamefont{Zhigadlo et~al.}()\citenamefont{Zhigadlo, Katrych,
  Bukowski, and Karpinski}}]{NDZhigadlo08060337}
\bibinfo{author}{\bibfnamefont{N.~D.} \bibnamefont{Zhigadlo}},
  \bibinfo{author}{\bibfnamefont{S.}~\bibnamefont{Katrych}},
  \bibinfo{author}{\bibfnamefont{Z.}~\bibnamefont{Bukowski}}, \bibnamefont{and}
  \bibinfo{author}{\bibfnamefont{J.}~\bibnamefont{Karpinski}},
  \bibinfo{journal}{arXiv:0806.0337 (unpublished)}.

\bibitem[{\citenamefont{Carrington et~al.}(1999)\citenamefont{Carrington,
  Giannetta, Kim, and Giapintzakis}}]{CarringtonGKG99}
\bibinfo{author}{\bibfnamefont{A.}~\bibnamefont{Carrington}},
  \bibinfo{author}{\bibfnamefont{R.~W.} \bibnamefont{Giannetta}},
  \bibinfo{author}{\bibfnamefont{J.~T.} \bibnamefont{Kim}}, \bibnamefont{and}
  \bibinfo{author}{\bibfnamefont{J.}~\bibnamefont{Giapintzakis}},
  \bibinfo{journal}{Phys. Rev. B} \textbf{\bibinfo{volume}{59}},
  \bibinfo{pages}{14173} (\bibinfo{year}{1999}).

\bibitem[{\citenamefont{Prozorov et~al.}(2000)\citenamefont{Prozorov,
  Giannetta, Carrington, and Araujo-Moreira}}]{ProzorovGCA00}
\bibinfo{author}{\bibfnamefont{R.}~\bibnamefont{Prozorov}},
  \bibinfo{author}{\bibfnamefont{R.~W.} \bibnamefont{Giannetta}},
  \bibinfo{author}{\bibfnamefont{A.}~\bibnamefont{Carrington}},
  \bibnamefont{and} \bibinfo{author}{\bibfnamefont{F.~M.}
  \bibnamefont{Araujo-Moreira}}, \bibinfo{journal}{Phys. Rev. B}
  \textbf{\bibinfo{volume}{62}}, \bibinfo{pages}{115} (\bibinfo{year}{2000}).

\bibitem[{\citenamefont{Fletcher et~al.}(2007)\citenamefont{Fletcher,
  Carrington, Diener, Rodiere, Brison, Prozorov, Olheiser, and
  Giannetta}}]{fletcher2007}
\bibinfo{author}{\bibfnamefont{J.~D.} \bibnamefont{Fletcher}},
  \bibinfo{author}{\bibfnamefont{A.}~\bibnamefont{Carrington}},
  \bibinfo{author}{\bibfnamefont{P.}~\bibnamefont{Diener}},
  \bibinfo{author}{\bibfnamefont{P.}~\bibnamefont{Rodiere}},
  \bibinfo{author}{\bibfnamefont{J.~P.} \bibnamefont{Brison}},
  \bibinfo{author}{\bibfnamefont{R.}~\bibnamefont{Prozorov}},
  \bibinfo{author}{\bibfnamefont{T.}~\bibnamefont{Olheiser}}, \bibnamefont{and}
  \bibinfo{author}{\bibfnamefont{R.~W.} \bibnamefont{Giannetta}},
  \bibinfo{journal}{Phys. Rev. Lett.} \textbf{\bibinfo{volume}{98}},
  \bibinfo{pages}{057003} (\bibinfo{year}{2007}).

\bibitem[{\citenamefont{Manzano et~al.}(2002)\citenamefont{Manzano, Carrington,
  Hussey, Lee, Yamamoto, and Tajima}}]{ManzanoCHLYT02}
\bibinfo{author}{\bibfnamefont{F.}~\bibnamefont{Manzano}},
  \bibinfo{author}{\bibfnamefont{A.}~\bibnamefont{Carrington}},
  \bibinfo{author}{\bibfnamefont{N.~E.} \bibnamefont{Hussey}},
  \bibinfo{author}{\bibfnamefont{S.}~\bibnamefont{Lee}},
  \bibinfo{author}{\bibfnamefont{A.}~\bibnamefont{Yamamoto}}, \bibnamefont{and}
  \bibinfo{author}{\bibfnamefont{S.}~\bibnamefont{Tajima}},
  \bibinfo{journal}{Phys. Rev. Lett.} \textbf{\bibinfo{volume}{88}},
  \bibinfo{pages}{047002} (\bibinfo{year}{2002}).

\bibitem[{\citenamefont{Hirschfeld and Goldenfeld}(1993)}]{HirschfeldG93}
\bibinfo{author}{\bibfnamefont{P.~J.} \bibnamefont{Hirschfeld}}
  \bibnamefont{and}
  \bibinfo{author}{\bibfnamefont{N.}~\bibnamefont{Goldenfeld}},
  \bibinfo{journal}{Phys. Rev. B} \textbf{\bibinfo{volume}{48}},
  \bibinfo{pages}{4219} (\bibinfo{year}{1993}).

\bibitem[{\citenamefont{Cooper}(1996)}]{Cooper96}
\bibinfo{author}{\bibfnamefont{J.~R.} \bibnamefont{Cooper}},
  \bibinfo{journal}{Phys. Rev. B} \textbf{\bibinfo{volume}{54}},
  \bibinfo{pages}{R3753} (\bibinfo{year}{1996}).

\bibitem[{\citenamefont{Liu et~al.}()\citenamefont{Liu, Kondo, Tillman, Gordon,
  Samolyuk, Lee, Martin, McChesney, Bud'ko, Tanatar, Rotenberg, Canfield,
  Prozorov, Harmon, and Kaminski}}]{CLiu08062147}
\bibinfo{author}{\bibfnamefont{C.}~\bibnamefont{Liu}},
  \bibinfo{author}{\bibfnamefont{T.}~\bibnamefont{Kondo}},
  \bibinfo{author}{\bibfnamefont{M.~E.} \bibnamefont{Tillman}},
  \bibinfo{author}{\bibfnamefont{R.}~\bibnamefont{Gordon}},
  \bibinfo{author}{\bibfnamefont{G.~D.} \bibnamefont{Samolyuk}},
  \bibinfo{author}{\bibfnamefont{Y.}~\bibnamefont{Lee}},
  \bibinfo{author}{\bibfnamefont{C.}~\bibnamefont{Martin}},
  \bibinfo{author}{\bibfnamefont{J.~L.} \bibnamefont{McChesney}},
  \bibinfo{author}{\bibfnamefont{S.}~\bibnamefont{Bud'ko}},
  \bibinfo{author}{\bibfnamefont{M.~A.} \bibnamefont{Tanatar}},
  \bibinfo{author}{\bibfnamefont{E.}~\bibnamefont{Rotenberg}},
  \bibinfo{author}{\bibfnamefont{P.~C.} \bibnamefont{Canfield}},
  \bibinfo{author}{\bibfnamefont{R.}~\bibnamefont{Prozorov}},
  \bibinfo{author}{\bibfnamefont{B.~N.} \bibnamefont{Harmon}},
  \bibnamefont{and} \bibinfo{author}{\bibfnamefont{A.}~\bibnamefont{Kaminski}},
  \bibinfo{journal}{arXiv:0806.2147 (unpublished)}.

\bibitem[{\citenamefont{Khasanov et~al.}(2008)\citenamefont{Khasanov, Luetkens,
  Amato, Klauss, Ren, Yang, Lu, and Zhao}}]{Khasanov08051923}
\bibinfo{author}{\bibfnamefont{R.}~\bibnamefont{Khasanov}},
  \bibinfo{author}{\bibfnamefont{H.}~\bibnamefont{Luetkens}},
  \bibinfo{author}{\bibfnamefont{A.}~\bibnamefont{Amato}},
  \bibinfo{author}{\bibfnamefont{H.-H.} \bibnamefont{Klauss}},
  \bibinfo{author}{\bibfnamefont{Z.}~\bibnamefont{Ren}},
  \bibinfo{author}{\bibfnamefont{J.}~\bibnamefont{Yang}},
  \bibinfo{author}{\bibfnamefont{W.}~\bibnamefont{Lu}}, \bibnamefont{and}
  \bibinfo{author}{\bibfnamefont{Z.-X.} \bibnamefont{Zhao}},
  \bibinfo{journal}{arXiv:0805.1923 (unpublished)}.


\bibitem{lam0note}  Note that if $\lambda(0)$ is assumed to be significantly below 2000\,\AA~the
calculated superfluid density would have have negative curvature over an large range of temperature which would be
highly unusual and is not compatible with this two gap model.


\bibitem[{\citenamefont{Fletcher et~al.}(2005)\citenamefont{Fletcher,
  Carrington, Taylor, Kazakov, and Karpinski}}]{Fletcher2005}
\bibinfo{author}{\bibfnamefont{J.}~\bibnamefont{Fletcher}},
  \bibinfo{author}{\bibfnamefont{A.}~\bibnamefont{Carrington}},
  \bibinfo{author}{\bibfnamefont{O.}~\bibnamefont{Taylor}},
  \bibinfo{author}{\bibfnamefont{S.}~\bibnamefont{Kazakov}}, \bibnamefont{and}
  \bibinfo{author}{\bibfnamefont{J.}~\bibnamefont{Karpinski}},
  \bibinfo{journal}{Phys. Rev. Lett.} \textbf{\bibinfo{volume}{95}},
  \bibinfo{pages}{097005} (\bibinfo{year}{2005}).

\end{thebibliography}
%\bibliographystyle{apsrevNOETAL}

\end{document}